\documentclass[aip,pop,amsmath,amssymb,reprint]{revtex4-1}

\pdfoutput=1
\usepackage{color}
\usepackage{xcolor}
\usepackage{graphicx}
\usepackage{bm}
\begin{document}
\preprint{AIP/123-QED}

\title[]{Turbulent electromagnetic fields at sub-proton scales: two-fluid and 
full-kinetic plasma simulations}

\author{C.A. Gonz\'alez}
 \email{caangonzalez@df.uba.ar}
\affiliation{Departamento de F\'isica, Facultad de Ciencias Exactas y Naturales, Universidad de Buenos Aires and IFIBA, CONICET, Ciudad universitaria, 1428 Buenos Aires, Argentina.}

\author{T. Parashar}
\
\affiliation{Bartol Research Institute and Department of Physics and Astronomy, University of
Delaware, Newark, Delaware, USA}

\author{D. Gomez}
\
\affiliation{Instituto de Astronomía y F\'isica del Espacio, CONICET-UBA, Ciudad Universitaria, 1428 Buenos Aires, Argentina}
\author{W.H. Matthaeus}
\
\affiliation{Bartol Research Institute and Department of Physics and Astronomy, 
University of Delaware, Newark, Delaware, USA}

\author{P. Dmitruk}
\
\affiliation{Departamento de F\'isica, Facultad de Ciencias Exactas y Naturales, Universidad de Buenos Aires and IFIBA, CONICET, Ciudad universitaria, 1428 Buenos Aires, Argentina.}


\begin{abstract}
Plasma dynamics is a multi-scale problem that involves many spatial and temporal
scales. Turbulence connects the disparate scales in this system through a cascade
that is established by nonlinear ineteractions. Most astrophysical plasma systems
are weakly collisional, making a fully kinetic Vlasov
description of the system essential. Use of reduced models to study such systems 
is computationally desirable but careful benchmarking of physics in different 
models is needed. We perform one such comparison here between fully kinetic 
Particle-In-Cell (PIC) model and a two-fluid model that includes Hall physics 
and electron inertia, with a particular focus on the sub-proton scale electric 
field. We show that in general the two fluid model captures large scale dynamics 
reasonably well. At smaller scales the Hall physics is also captured reasonably 
well by the fluid code but electron features show departures from the
fully kinetic model. Implications for use of such fluid models are discussed.

\end{abstract}

\maketitle

%

\section{\label{sec:level1}INTRODUCTION:}

Turbulence is generally defined as an ensemble of broadband fluctuations
that arise from the nonlinear interaction among many degrees of freedom, which 
involves the energy transfer between scales. In hydrodynamic 
flows, energy is transferred or cascaded from large to progressively smaller  
scales where it is finally dissipated. The corresponding energy spectrum displays a 
Kolmogorov power-law\cite{k41,coleman68,Golstein1995} that represents  
scale invariance in a given range. The large-scale behavior of turbulent plasmas, 
traditionally described within the framework of magnetohydrodynamics, is also characterized by a 
direct energy cascade. The nature of this cascade and the
corresponding effect on plasma heating are important problems in 
solar and space physics.  The dynamics of these plasmas are complex, with many temporal and spatial scales simultaneously involved. For instance, the energy transfer processes that occur at proton and electron inertial scales in collisionless
plasmas, involve coherent features such as tube-like structures \cite{Bruno2007,Alexandrova2012a,Perri2012,Perrone2016} and intermittent
events that emerge \cite{Bruno2001,Greco2012,Zhadankin2012,Greco2008} through the coupling 
between scales. These process are also related to the production of high energetic particle populations\cite{Tessein2013,Tessein2015,PD1,Gonzalez2017}. 

\textit{In-situ} observation of the solar wind has allowed research to go deeper into the 
physics of plasma turbulence. In spite of the progressively higher resolution 
cadence of recent space missions, there is still a debate about the characteristics of the turbulence
near the dissipation range. The solar wind displays a Kolmogorov-like power-law at inertial range . However, when approaching the
proton inertial scale, the magnetic energy spectrum displays a break \cite{Leamon98,SmithEA2006} toward a somewhat steeper power law. The spectral 
index at proton scales can vary depending on the physical processes acting on the
flow, such as magnetic 
reconnection\cite{cerri2018dual,vech2018magnetic,loureiro2018turbulence}, damping 
of kinetic Alfv\'en waves (KAWs), 
anisotropies in the proton distribution function or differential particle heating.
The details of the energy cascade beyond the electron inertial range is even less
understood, although observations show a much steeper magnetic energy spectrum. 
There is still no agreement in the
plasma physics community about  the dominant dissipation mechanisms at these small
scales\cite{Alexandrova2012,parashar2015_tdc}. For instance, finite electron 
larmor effects\cite{Alexandrova2009,Sahraoui2013}, electron-cyclotron resonance, 
magnetosonic, whistler and/or KAW mode turbulence\cite{Sahraoui2010,Salem2012,
Saito2010,Narita2010,Podesta2011,Hughes2017,Gros2018,Cerri2016_electric} are some 
of the likely  candidates to mediate the energy transfer and dissipation at electron scales. 

Solar wind observation evidence that the magnetic and electric field 
spectra are well correlated on the inertial range, and this correlation is in 
agreement with  the scenario of Alfv\'enic turbulence\cite{Mozer2013}. It is well
stablished that there is a transition from MHD turbulence to kinetic turbulence, 
where the electric and magnetic field spectra departs each other at around proton
scales, with a relatively more intense electric field and with a 
shallower slope \cite{Bale2005,Matteini2017}.
\textit{In-situ} measurements\cite{Salem2012,Sahraoui2010,Narita2010} are in
agreement with theoretical 
predictions\cite{Schekochilin2009} about the kinetic Alfv\'en and whistler regimes 
at proton and electron inertial scales, where a parallel electric field component is generated. The importance 
of the parallel electric field fluctuations is due to the relevance of Landau 
damping and the wave-particle interaction that might become significant at kinetic 
scales in turbulent plasmas\cite{Quataert1998}.

From a theoretical point of view, the  dynamics of plasmas and 
their self-consistent electric and magnetic fields can alternatively be described 
by multi-fluid or kinetic descriptions. According to the more fundamental kinetic 
approach, the dynamics is described by the Vlasov equation for each plasma 
species, coupled with Maxwell's equations for the electric and magnetic fields. The relatively simpler 
multi-fluid models are derived from the moments of the Vlasov equation and the plasma dynamics 
for each species is described by fluid quantities such as its density, velocity field and pressure. 
There are several different numerical methods based on fluid-like descriptions, 
depending on the physical processes involved. In the case of magnetohydrodynamics
(MHD) the dynamics is described by a single fluid. Other fluid based methods
include, but are not limited to Hall-MHD \cite{Ghosh97}, Electron 
MHD\cite{das2000,shaikh2009}, Electron Reduced MHD \cite{Schekochilin2009},
Electron inertial Hall-MHD\cite{andres2014}, Hall-Finite Larmor Radius
MHD\cite{yakima96,Ghosh96} and Landau-Fluid \cite{Hammet90,passot2004}.

There are also numerical methods based on a full kinetic description\cite{vencels2016,JUNO2018}, often
constrained by computational limitations. One possibility is to integrate the 
Vlasov equation for the dynamics of each species, coupled to Maxwell's equations 
for the electric and magnetic fields \cite{JUNO2018}. There is also
the case of full kinetic 
particle-in-cell method (PIC) where the plasma species, both ions and electrons,
are treated as computational macro-particle which represents a number of 
real particles with similar physical properties and in a close 
region of the phase space\cite{birdsall,hokney}. Besides fluid and kinetic methods,
there are other hybrid methods that typically consider the 
electrons as a fluid and kinetic protons or variations of that.\cite{VALENTINI2007}

In this paper we perform a comparison between a two-fluid and a full PIC 
simulation where the dynamics of both protons and electrons and the corresponding 
inertial scales have been taken into account. The comparison made in this work can be considered 
within the framework of the \textit{``Turbulent dissipation challenge''} \cite{parashar2015_tdc}, 
in the spirit of comparing different 
simulation models under the same initial conditions with similar physical
and numerical parameters despite the limitations of each numerical and theoretical
framework. Several attempts to compare different numerical methods have been carried out in
recent years\cite{Cerri2017,Groselj2017,Stanier2017,Franci2017,Pezzi2017} and it has
resulted in a very useful and interesting approach to explore different physical mechanisms in plasma 
turbulence.

While in the two-fluid simulation the system is driven by the dynamics of the flow,  
the PIC simulation allows for the interaction between the field and individual macroparticles, thus enabling  new channels for the exchange of 
energy \cite{Klein2016,Yang2017a,Yang2017,howes_klein_li_2017,Chasapis2018}. 
Also, many physical questions must be addressed in order to shed light on the importance
of intermittency, high and low-frecuency wave phenomena and particle
heating on dissipation processes in collisionless plasmas. The goal of 
this paper is to delve more deeply
into the relationship between large-scale and 
kinetic scale fluctuations, noting both similarities and differences in two very 
different plasma  models. In particular, we explore contributions to  the 
electric field at sub-proton scales where kinetic contributions may strongly 
influence dissipation and charged particle energization.

This paper has the following organization: in section II. we present the 
kinetic and the two-fluid  models. In section III we describe
the initialization setup for both simulations and introduce the relevant 
plasma parameters. In section IV. we present the results of the
simulation. First, we compare  important macroscopic quantities for both simulations, like flow energy, current 
density and the magnetic field. After that, we study the electric field at different spatial 
scales for both simulations, comparing the various terms of the generalized Ohm's law with the aim 
to unveil which terms are dominant at sub-proton scales. 
Conclusions and directions for future work are established in Section V.

\section{\label{sec:level2}MODELS:}

This section describes the  
models employed for this study. We consider a
collisionless approximation where the 
dynamics of each species $s$ is governed by the kinetic Vlasov equation:

\begin{equation}
\frac{\partial f_s}{\partial t} + \textbf{v} \cdot \nabla f_s + \frac{q_s}{m_s}\left( \textbf{E} + \frac{\textbf{v} \times \textbf{B}}{c} \right) \cdot \frac{\partial f_s}{\partial  \textbf{v}} = 0 
\end{equation}

The evolution of the distribution function $f_s(\textbf{r},\textbf{v},t)$ for
the plasma species $s$ with a given charge $q_s$ and mass $m_s$ depends on the 
external and self-consistent electromagnetic fields 
$\textbf{E}$ and $\textbf{B}$,
that follow Maxwell's equations:

\begin{equation}
\frac{\partial \textbf{E}}{\partial t} = c \nabla \times \textbf{B} - 4 \pi \textbf{J}, \  \  \  \  \  \  \ \ \frac{\partial \textbf{B}}{\partial t} = - c \nabla \times \textbf{E}
\end{equation}
\begin{equation}
\nabla \cdot \textbf{E} = 4 \pi \rho, \  \  \  \  \  \  \ \ \nabla \cdot \textbf{B} = 0
\end{equation}

\noindent The sources of Equations (2) and (3) (i.e. the charge density $\rho$ and 
the electric current density $\textbf{J}$) are obtained directly 
from the distribution function of all plasma species 
through: $\textbf{J} = \sum_s {q_s} \int{\textbf{v}f_sd^3\textbf{r}}$ 
and $\rho = \sum_s {q_s} \int{f_sd^3\textbf{r}}$. 

On the other hand, the fluid model adopted for this paper 
is a two-fluid MHD, that extends MHD 
to include the dynamics of both protons and electrons. The model 
contains the momentum equation for the both species ($s = e,i$):

\begin{equation}
m_s n\frac{d\textbf{u}_s}{dt} = q_s n\left( \textbf{E} + 
\frac{\textbf{u}_s \times \textbf{B}}{c} \right) - \nabla p_s + 
\mu_s\nabla^2\textbf{u}_s \pm \textbf{R}
\end{equation}

\noindent With the total derivative  $d/dt = \partial /\partial t
+ \textbf{u}_{s} \cdot \nabla$, the pressure $p_s$ , $\textbf{u}_s$ is
the velocity field and $\mu_{s}$ is the 
viscosity for a given plasma species. \textbf{R} is the rate of
momentum exchanged between protons 
and electron through colissions and it is assumed to be proportional
to the relative speed between plasma species $R = -nm_i\nu_{ie}(\textbf{u}_i-\textbf{u}_e)$,
where $\nu_{ie}$ is the collisional frequency of an ion against
electrons. Note that although in the fluid model the
collisions between plasma species must be taken into 
account since it determines the smallest timescale in the
model, we are interested in studying and comparing colissionless
phenomena at relatively larger scales. Therefore, we restrict
ourselves to examination of 
quantities that relate to 
spatial and temporal scales common to both models. 

We are interested in studying the non-relativistic limit, and so
the displacement current can be neglected. Then the current density 
can be written as follows:

\begin{equation}
 \textbf{J} = \frac{c}{4\pi}\nabla \times \textbf{B} = en (\textbf{u}_i - \textbf{u}_e) 
\end{equation}
\begin{equation}
n_i = n_e = n \ \ \ \ \nabla \cdot \textbf{u}_{i,e} = 0
\end{equation}

\noindent We assume quasi-neutrality and incompressibility
 for both proton and electron fluids (Equations. (6)). We refer the reader 
to Andres et al. (2014) paper for a complete description of the 
two-fluid model. This fluid model, where the electron mass is not neglected,
has been called EIHMHD, for electron inertia Hall MHD \cite{andres2014}.

An expression for the electric field emerges from 
the electron momentum equation, leading to the 
generalized Ohm's law\cite{vasylinas2005,Krishnaswami2017,VALENTINI2007}:

\begin{center}
\begin{equation}
\small{\textbf{E} =  - \textbf{U} \times \textbf{B} }
\small{+\frac{1}{n}\textbf{J} \times \textbf{B} -
\frac{1}{n} \left(\nabla p_e + \nabla \cdot \Pi_e\right) -}
\small{\frac{d_e^2}{n}\frac{\partial \textbf{J}}{\partial t}+}
\small{\eta \textbf{J} }
\end{equation}
\end{center}

With $\textbf{U} = \frac{(m_i \textbf{u}_i + m_e \textbf{u}_e)}{(m_i+m_e)}$ 
the bulk velocity,
$p = \frac{1}{3}P_{ii}$ and $\Pi_{ij} = P_{ij} - p \delta_{ij}$,
one may conveniently 
decompose the electron pressure tensor into 
isotropic and deviatoric parts. 
Equation (7) is a normalized version of the electric field
in units of the proton inertial length $d_{i} = c/\omega_{pi}$ and
Alfv\'en velocity $v_A = B_0/(4\pi n m_i)^{1/2}$, appropriate  
for a 
magnetized plasma. The RHS terms 
are the induction (ideal MHD) 
term, the Hall term 
(related to differential flow of ions and electrons), which is 
important at scales $<d_i$. The
isotropic and anisotropic 
electron pressure terms are 
also important at scales smaller than  $d_i$.
The electron inertial term 
becomes relevant at electron inertial 
length  $d_e = c/\omega_{pe} = \sqrt{m_e/m_i} \ d_i$. 
Those terms deal with frequencies much lower 
than the electron plasma frequency and could mediate some phenomena that
occur on sub-proton scales\cite{che2018}. The last
term is the contribution to the electric field that results from 
collisional resistivity.
Ohm's law is valid for a system size larger than the Debye length
($L \gg\lambda_D$) and for plasma frequencies slower than 
the electron plasma frequency ($\omega \ll \omega_{pe}$). At scales smaller than $ \lambda_D$
and frequencies greater than $\omega_{pe}$, the quasi-neutrality condition becomes
invalid.  The validity of this equation is also constrained to a weakly magnetized plasma ($\omega_{ce} \ll \omega_{pe}$) with $\omega_{ce}=eB/m_e c$ the
electron cyclotron frequency around the mean magnetic field. 
It is extremely difficult to simulate such values in fully kinetic models as the
simulations become prohibitively expensive in this regime. Typical fully kinetic
simulations have $\omega_{ce}/\omega_{pe} \sim 3$ or so.

\section{\label{sec:level3}SIMULATION SETUP AND THE INITIAL
CONDITIONS}

The simulations presented in this paper are in 2.5D dimensions 
(two-dimensional dependence  and all three
components of the field vectors). We use periodic boundary conditions with an
out-of-plane mean magnetic field $B_0$ for both kinetic and
two-fluid simulations. The kinetic simulation is performed with
the electromagnetic PIC code P3D \cite{Zeiler2002}. This is a 
finite difference particle code, where the particles are advanced using the Boris scheme and the 
Maxwell's equations are advanced an explicit trapezoidal leapfrog
algorithm. The two-fluid simulation is done employing a Fourier 
pseudospectral code with a second-order Runge-Kutta time 
integration scheme.
The system size in the PIC simulation has $L=L_x=L_y=149.6 d_i$
with $N_x=N_y=4096$ number of grid points. Here, we study plasma 
with initial ion and electron temperature equilibrium ($T_i/T_e=1$), with
a mass ratio $m_e/m_i=0.04$ and $\beta_e=\beta_i = 0.6$. The 
proton scale is at $k_{d_{i}}  \sim 23.8 k_{box}$, the electron 
scale appears at $k_{d_{e}} \sim 120 k_{box}$ and the Debye length is
at $k_{\lambda_D}=652 k_{box}$, where $k_{box}$ is the wave-number corresponding
to the largest wavelength that can fit in the box in the PIC simulation.
With the above mentioned parameters we capture the MHD-like 
behavior that results self-consistently in kinetic simulations\cite{Parashar2015}.

The system size must be large enough to represent the 
energy evolution expected in the MHD simulation and the 
development of a broad inertial range in the spectrum. The 
PIC simulations are also susceptible to noise effects due to 
the statistics in the number of macro particles per grid cell (ppc) used
in the simulations. The number of ppc 
is important for the fields accuracy at 
Debye length where the thermalization may have a dramatic effect
on the features of kinetic simulations\cite{Turner2006,Haggerty2017}. We
use $3200 \ ppc$ in order to reduce those effects on
the Debye scale and compare the entire evolution
with the two-fluid simulation where the Debye length is not 
resolved.

The two-fluid simulation is in a normalized square box 
of length $L=2\pi L_0$, where $L_0$ is the characteristic length 
(also called energy containing scale) defined 
as $L_0 = \int{(E(k)/k) \ dk}/\int{E(k) \ dk}$ with $E(k)$
the energy spectral density at wavenumber $k$. 
To suppress the aliasing effects, the code uses a maximum wavenumber
$k_{max}=N/3$ where $N$ is the Fourier modes resolution in the 
simulation. 
For this simulation, we used a resolution of $2048$ Fourier modes,
with equal viscosity and resistivity ($\eta=\nu=7.5 \times 10^{-6}$). With the
same mass ratio used in PIC simulation ($m_e/m_i=0.04$), the 
corresponding proton scale is at $k_{d_{i}}  \sim 25$ in the 
simulation, the electron scale appears at $k_{d_{i}} \approx 125$
and the dissipation scale is $k_{diss}=356$. $k_{diss}= \langle j^2 + w^2\rangle^{1/4}/\sqrt{\mu}$. is the viscous dissipation scale that depends on the energy dissipation
rate and the viscosity of the flow. That scale 
lacks physical sense for collisionless plasmas and for kinetic 
models where the energy dissipation is due to other phenomena. 

Our goal in the following sections is to compare the results arising 
from these two descriptions, restricting ourselves to the
scales that both models share in common.
We have done freely decaying turbulence simulations and the initial 
conditions are chosen such that the root mean square (rms)
of magnetic fluctuation  $\langle \textbf{b}\rangle/B_0 = 1/3.16$, and
the Alfv\'en ratio of fluctuating kinetic and magnetic energies 
is $E_v/E_b = 1$. We initially excite a shell in k-space 
with wavenumber $2\leq |{\bf k}| \leq 4$, with a specified spectral shape
and Gaussian random phases. 
\begin{figure}[h!]
\begin{center}
{\includegraphics[width = 0.6\textwidth]{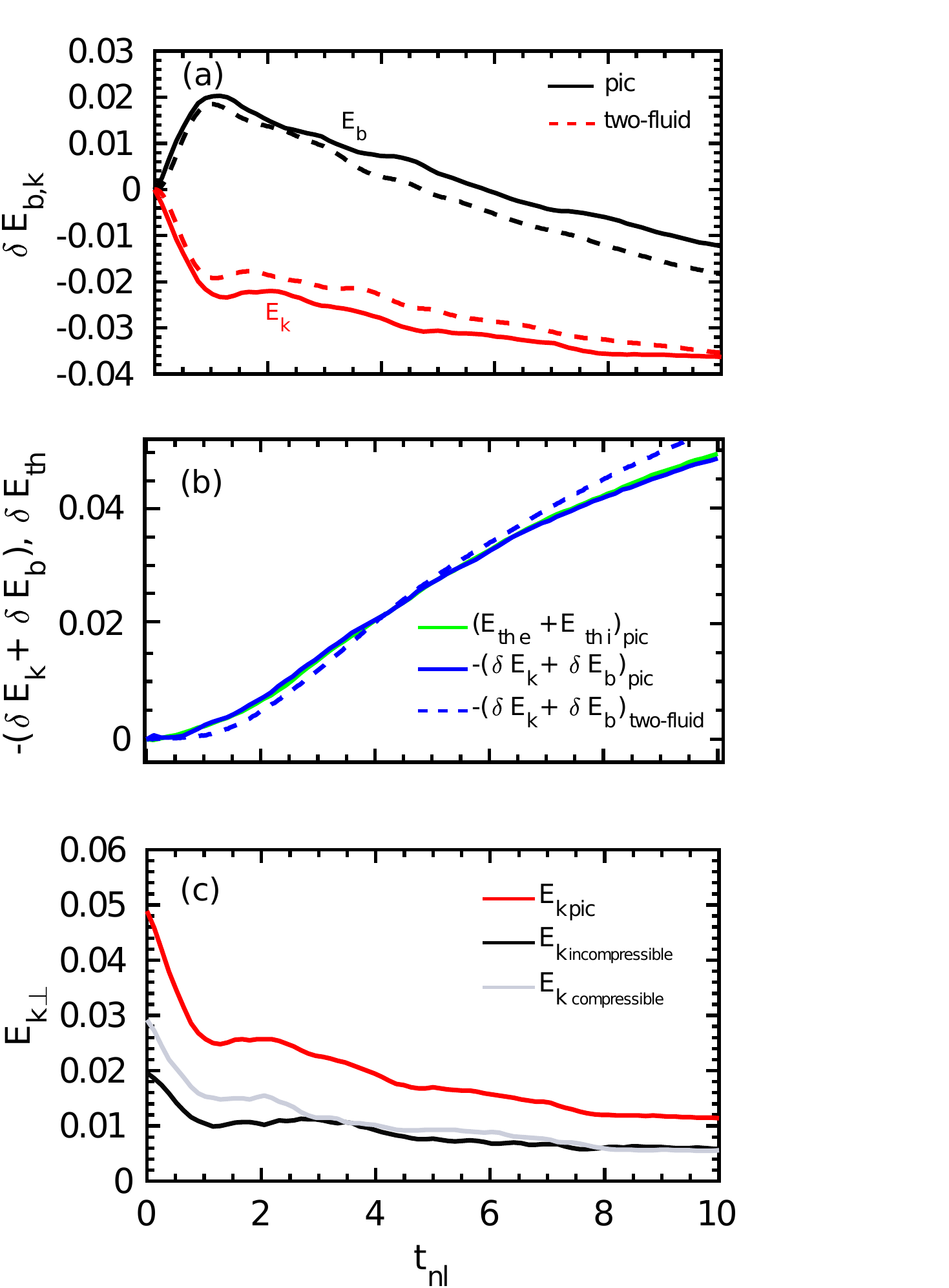}} 
\caption{[Top] (a). Time evolution of the variation in the magnetic
energy (black), kinetic flow energy (red) for the 
full-kinetic simulation (solid lines) and for the two-fluid 
simulation (dashed line).  [Middle] (b). The Variation of the total thermal
energy (green) and the negative value of the flow energy dissipation for both 
simulations (blue). [Bottom] (c) incompressible (black), compressible 
(gray) and total (red) perpendicular kinetic energy for the PIC 
simulation.}
\end{center}
\label{as}
\end{figure}

As those codes represent different plasma models with different 
normalization scheme, in order to compare the results of kinetic 
and two-fluid simulation, we take as a common characteristic timescale the
nonlinear time (or eddy turnover time) defined 
as $t_{nl}= L/\delta u$, with $\delta u$ the (r.m.s) velocity
value and $L$ the system size for each simulation.

\section{\label{sec:level4}RESULTS:}

In Figure 1 we present some global quantities for both fluid and kinetic 
simulations. The upper panel describes the time evolution of the energy per 
unit mass for PIC (solid lines) and two-fluid (dashed lines) simulations. 
Each quantity is plotted as its departure from the corresponding 
initial value, e.g., $\delta E_b = E_b(t) - E_b(0)$. After an initial adjustment, 
all quantities decay in time. The kinetic (red lines) and 
magnetic energy (black lines) show similar behavior but the magnetic energy is 
smaller in the two-fluid case than in the kinetic simulation, while the 
kinetic energy is larger in the fluid simulation. A possible reason for this 
discrepancy could be the absence of viscous-like Pi-D 
interactions (\cite{Yang2017a,Yang2017}) in the fluid case, and the absence of 
resistive-like  behavior in the kinetic case.

\begin{figure}[h!]
\begin{center}
{\includegraphics[width = 0.48\textwidth]{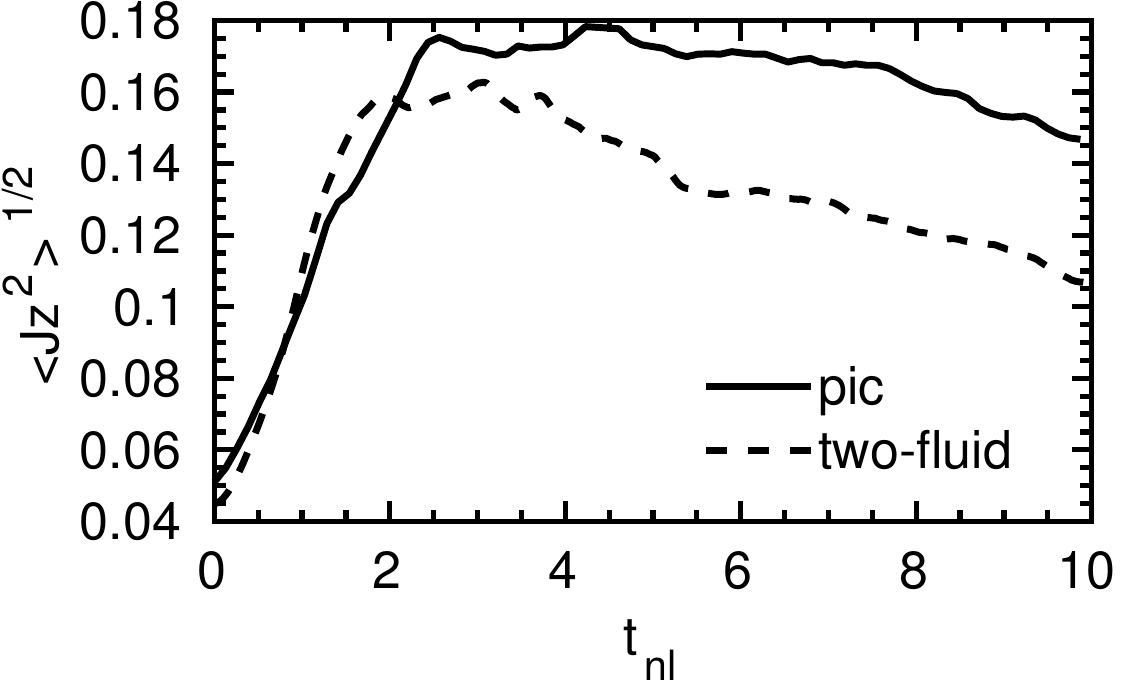}} 
\caption{Time history of the root mean square 
parallel electric current density ($J_z$) for the 
full-kinetic simulation (solid line) and r.m.s electric current 
density for the two-fluid simulation (dashed line)}
\end{center}
\label{as}
\end{figure}

The interaction of particles and fields in the PIC simulation 
enables channels that support dissipation processes. 
Interchange between flow and thermal energy is essential for physical dissipation, 
but the two-fluid (incompressible, isothermal) simulation lacks such couplings and 
therefore all the energy transported from the macro-scale to the smaller scales is
artificially removed by viscous and resistive effects. 
The total variation of the thermal energy 
($\delta {E_{th}}_i + \delta {E_{th}}_e$) in PIC, and the total variance of the flow
energy $-(\delta {E_{k}} + \delta {E_{b}})$ in both cases, are also shown in the middle
panel of Figure 1 (green and blue lines respectively). The changes $\delta E_{th}$ 
and $-(\delta E_k + \delta E_b)$ follow each other almost perfectly for the PIC
case, indicating extremely good energy conservation 
(see also \cite{parashar2018}). The change $-(\delta E_k + \delta E_b)$ for the
fluid run interestingly follows the PIC values very closely inspite of the
differences in quantitative behavior of individual magnetic and flow energies. This result suggests
that von Karman phenomenology, in which large scale fluctuations control the decay rate, is obtained for both of these cases. . 

In addition to thermal couplings, the PIC simulation also contains flow 
compressions that are clearly absent in the incompressible two-fluid model.
The extent of this difference is quantified using a Helmholtz decomposition of
the velocity field, separating the solenoidal and irrotational components. These
are equivalent to the incompressible 
( $\textbf{\^{u}}_i(k) = (\textbf{\^{I}} - \textbf{\^{k}}\textbf{\^{k}}) \ \textbf{u}(k)$ )
and the compressible part 
($\textbf{\^{u}}_c(k) = \textbf{u}(k) - \textbf{\^{u}}_i(k)$). The bottom panel of
Figure 1 shows the perpendicular component of the
incompressible kinetic energy (black line), the compressible kinetic
energy (gray line) and the total kinetic energy (red line) for the PIC simulation.
We observe that the compressible component is the dominant part of the 
kinetic flow energy in the system but both components become of the
same order of magntiude a few non-linear
times later. The compression of the flow allows a
channel where an interchange of energy between incompressible and 
compressible modes can occur. There is also an energy cascade
in each channel. Those effects seem to be very important at 
sub-proton scales where the compressive coherent structures play a
relevant role in the intermittency and the charged particle dynamics.

\begin{figure*}
\begin{center}
{\includegraphics[width = 0.485\textwidth]{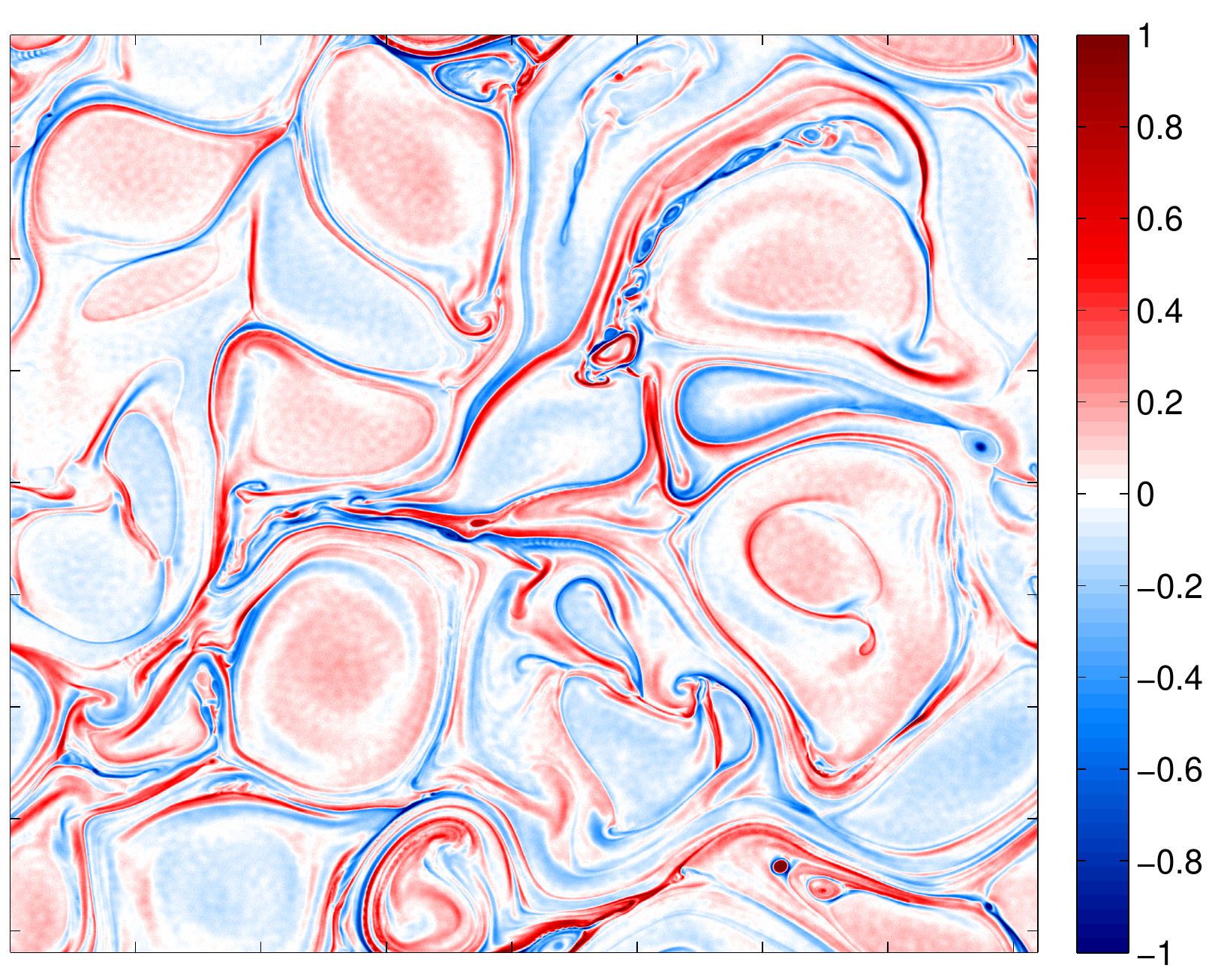}} 
{\includegraphics[width = 0.485\textwidth]{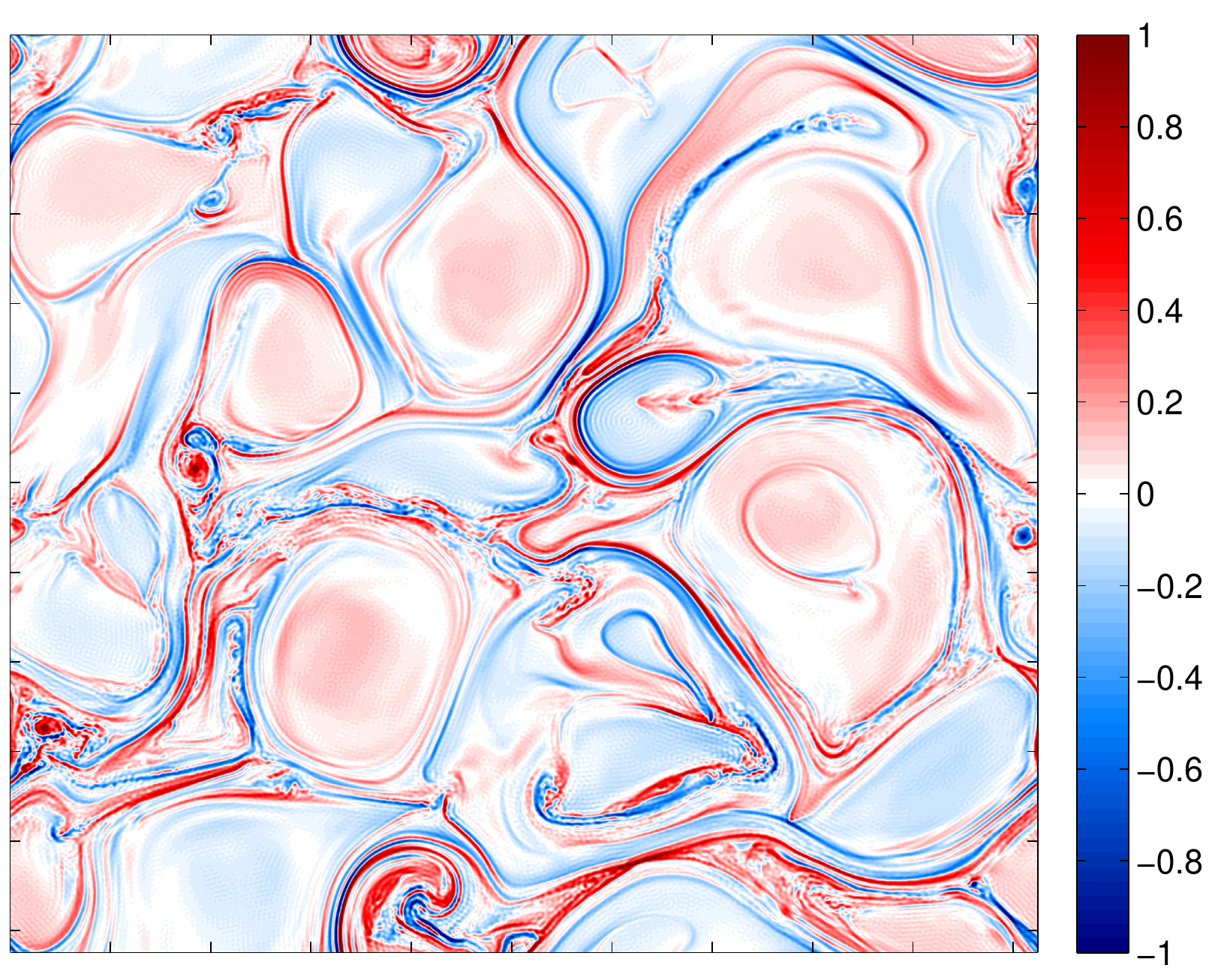}}
\caption {Contour plot of  the out-of-plane normalized 
electric current density ($J_z$) at the time of the maximum 
enstrophy in each simulation ($t \approx 3 t_{nl}$). [Left] snapshot 
of $J_z$ for the full-kinetic simulation. [Right] the same
for the two-fluid simulation.}
\end{center}
\end{figure*}

The root mean square value of the current density
in the system is a good indicator of dissipation 
in plasmas and the increase of this quantity
is a consequence of generation of many current sheets
mediated by magnetic reconnection. The time evolution of the r.m.s. of the
parallel (out of plane) 
current density $J_z$  is presented in Figure 2. 
The curves are 
roughly similar but the peak and late time vales of 
r.m.s current density 
remain systematically larger 
in the kinetic PIC 
simulation.  Once the 
peak is reached in Fig. 2,
at about $t= 2 t_{nl}$, some current sheets
become unstable and the plasma experiences 
a proliferation of secondary islands or 
plasmoids due to 
either turbulent cascade \cite{WanEA16}
or due to secondary instability 
\cite{Cerri2017a,loureiro2018turbulence}, or both.
The generation of a chain of magnetic islands, becomes possible, 
which in turn could be very important at the electron scales 
where particles can interact with these electromagnetic
structures. Particles with small gyroradious can 
easily violate the conservation of the adiabatic 
moments\cite{DahlinEA14,Dalena2012,leRouxEA2015,ZankEA14,DahlinEA17}
in such interactions.
The Figure also shows that the maximum of dissipation is reached early in 
the fluid simulation. This is a consequence of the fluid viscosity
and resistivity that dissipates the energy at a rate slightly faster than in the
kinetic simulation.

Cross-sections of the out-of-plane current density are shown 
in Figure 3, which depicts the normalized  $J_Z$ at $t = 3 t_{nl}$
for kinetic (Left tpanel) and fluid simulation (right panel). 
Large current sheets and macroscopic structures are notably similar
in these simulations and the signatures of magnetic 
reconnection are evident in many places in the simulation box.
However, the PIC simulation has a more refined small scale 
structure and some structures are quite different in spatial
positions and shape. At this point in the evolution, the system
has developed the conditions to the onset of secondary plasmoid 
formation, through either instability or turbulence cascade effects. 
As a result, individual small islands or  chains of small islands
are produced in the interior of large current sheets.

\begin{figure}[h!]
\begin{center}
{\includegraphics[width = 0.52\textwidth]{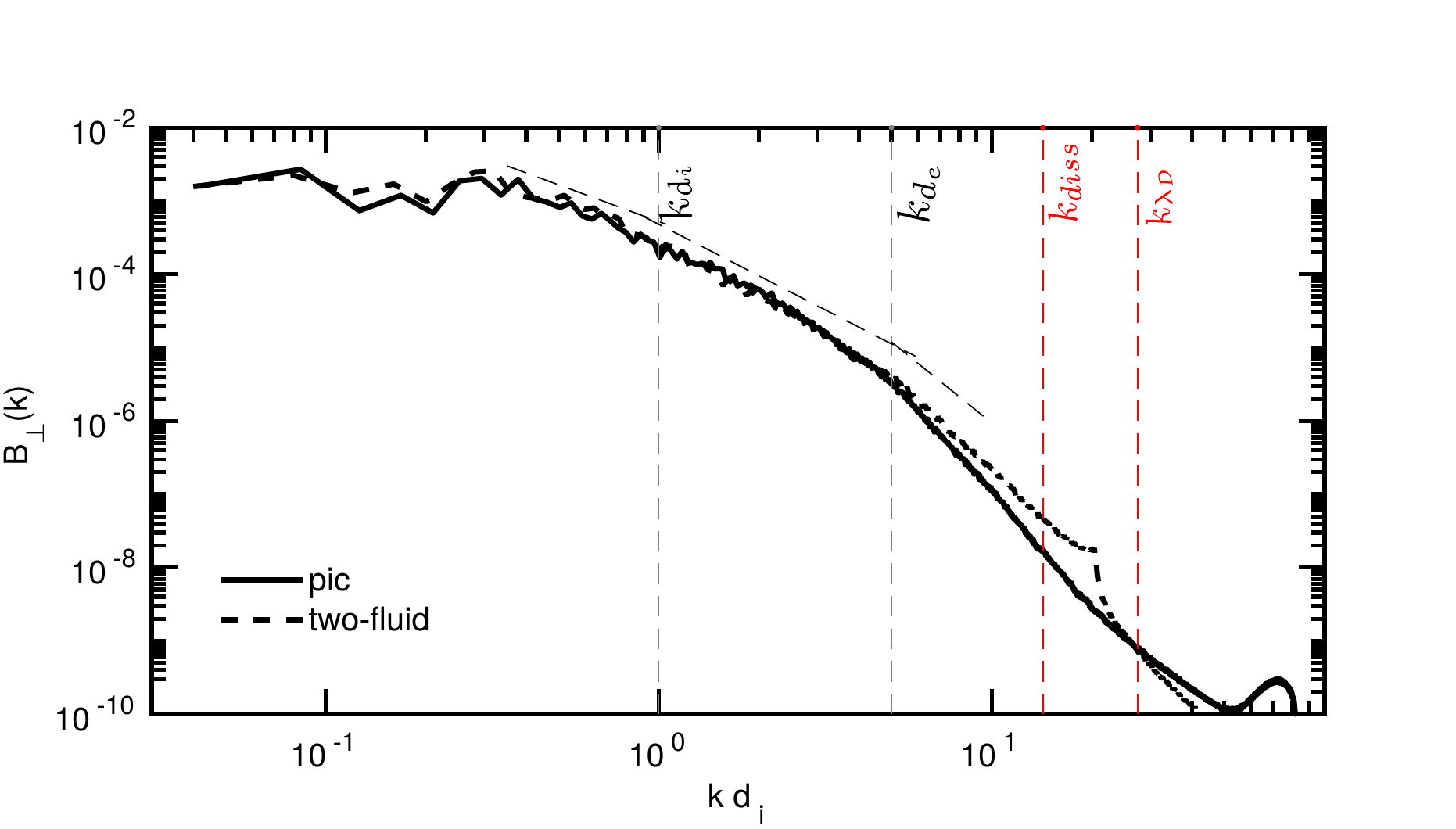}} 
\caption{Energy spectrum of the perpendicular components of the 
magnetic field at $t \approx 3 t_{nl}$ for PIC simulation (solid line)
and for the two-fluid simulation (dashed-lines). Vertical gray dashed 
lines represent the wavenumbers associated with the ion and electron inertial
lengths. Red dashed lines represent the smallest scale for 
each model, dissipation scale (fluid model) 
and the Debye length (PIC simulation).}
\end{center}
\end{figure}

The similarities in the magnetic structure in 
the simulations also carries over to a statistical description. 
In Figure 4, we show the perpendicular power spectra of the magnetic field at 
$t \approx 3 t_{nl}$ for the kinetic (solid line) and 
the two-fluid simulations (dashed line). 
From the figure one can  distinguish breaks in the magnetic spectrum at the 
particle inertial scales ($k_{\perp}d_i=1$ and $k_{\perp}d_i=5$ for ions and 
electrons respectively).

\begin{figure*}
\begin{center}
{\includegraphics[width = \textwidth]{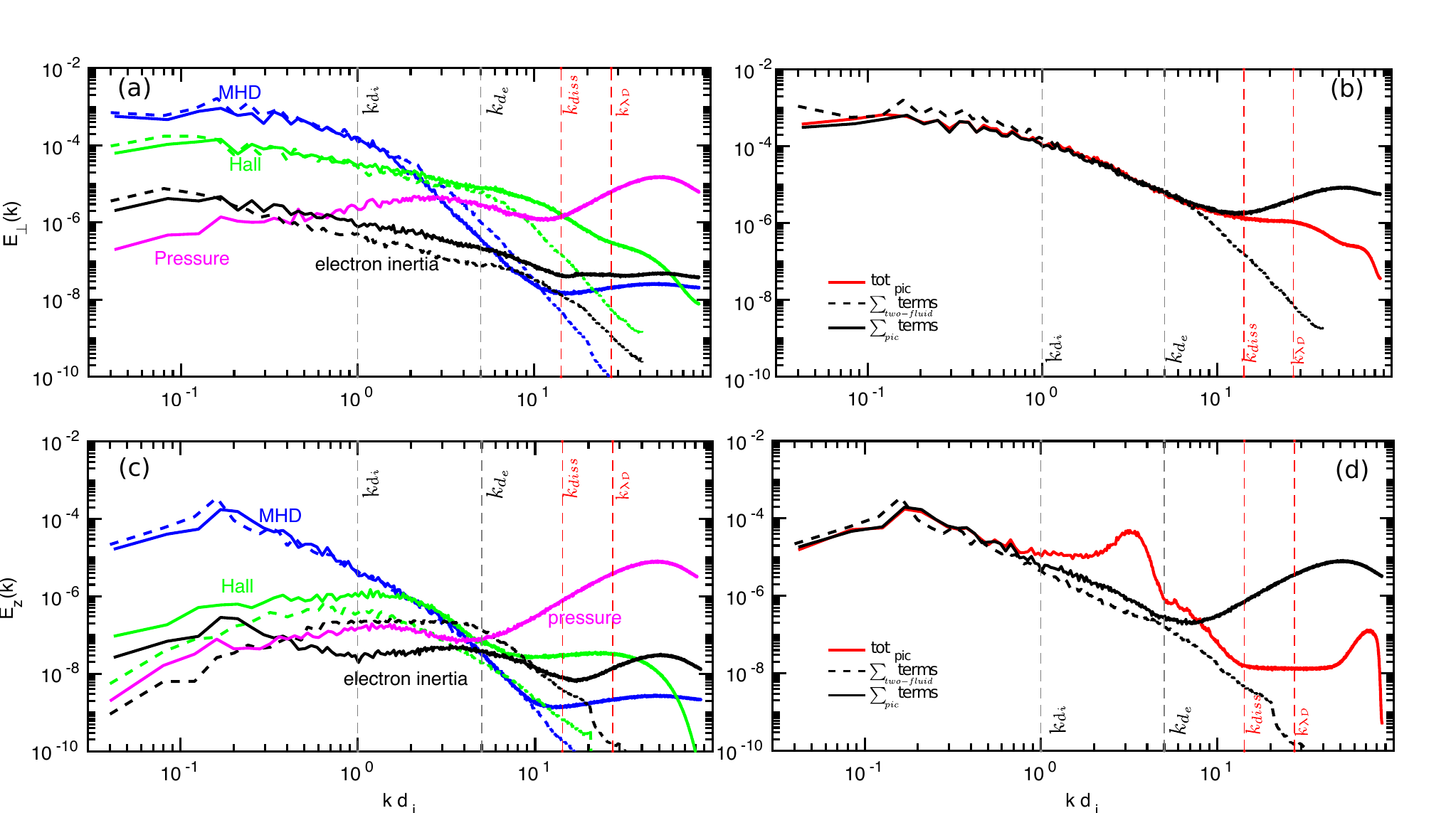}}
\caption{(\textit{Left-Top}.) (a). Power spectrum of the perpendicular
electric field for both simulations. The relevant electric field
terms for kinetic simulation (solid lines) and for two-fluid 
simulation (dashed lines). (\textit{Right-Top}.) (b) Contributions to the total perpendicular
electric field due to the terms in the generalized Ohm's law
for kinetic simulation and for the two-fluid simulation (black lines) and 
the total electric field in the PIC simulation by solving Maxwell 
equations (red line). (\textit{Left-Bottom}.) (c). Spectrum of the 
parallel component of the electric field for both simulations. The 
relevant electric field terms for kinetic simulation and for two-fluid 
simulation with the same line style used in the top panels. (\textit{Right-
Bottom}.) (d). Spectrum of the total parallel electric field from the generalized Ohms 
law for both types of simulations, and from direct output of the PIC simulation.  (same line style employed in top panels). Vertical gray dashed
lines represent the wavenumbers associated with the ion and electron
inertial lengths. Red dashed lines represent the
dissipation scale and the Debye length, respectively.}
\end{center}
\end{figure*}

Once the turbulence is fully developed and the maximum of 
dissipation is reached, the simulations exhibit 
the same qualitative behavior at the large scales. 
The spectral slope shows a Kolmogorov-like power-law in the
inertial range for both PIC and two-fluid simulations. A broader 
inertial range may be obtained in the kinetic simulation 
only with an intensive use of computational resources. For 
two-fluid simulations however, we can reproduce powerlaw-like features 
down to the ion and electron scales, at a remarkably lower computational cost.

Figure 4 also shows that near the proton scales the spectra from both 
simulations are in good qualitative agreement. The spectral 
index in this range of scales has a value between $k_\perp^{-7/3}$ 
and $k_\perp^{-8/3}$. A universal value of the spectral index
at sub-proton scale has not been established and a different value
have been observed in numerical simulations with different physical assumptions
\cite{Biskamp99,Howes2008,shaikh2009,Passot2014,Wan2012,Franci2016}.
Variation of the sub-proton range slope is also seen in solar wind observations, 
usually between $-1.75$ to $-3.75$\cite{Leamon98, SmithEA2006,Salem2012,Sahraoui2010}. There is some evidence for the dependence of the spectral index 
on compressibility (different value for fast and slow wind and a dependence
with the plasma beta\cite{Bruno2014,Franci2016,Cerri2017a,Groselj2017}), and on cascade rate 
\cite{SmithEA2006} (steeper spectra for stronger cascade). 

The electron scale spectrum is
not at all similar in the two 
simulations, as can be seen in Figure 4.
It is worth mentioning that there is no 
consensus in the community 
concerning the functional
form of the magnetic spectrum 
at electron 
scales\cite{Alexandrova2009,Alexandrova2012a,Sahraoui2010,Sahraoui2013}. However
the finite electron gyroradii effect must play
an important role and those effects are not retained in the
two-fluid simulation.

The large-scale structures and the resulting small-scale 
structures produced by the shrinking of sheets into a smaller-scale 
magnetic structures is a very important feature in plasma 
turbulence. The relevance of the coherent structures for particle 
energization is driven by the electric field that has the 
capacity to do work on the particles. In order to shed light on 
the nature of the electric field across scales, 
we study the contributions to  the generalized
Ohm's law terms for both types of numerical simulations. Simple scaling analysis can be used to understand the scales at which different terms in Ohm's law become important. However, the exact breakdown of importance of all the terms in parallel and perpendicular components, and across different models is a non-trivial task. We show this breakdown of all the terms in figure 5.

Figure 5 (a) compares contributions 
to the perpendicular component of the electric field spectrum,
from the generalized Ohm's law and from the fluid and kinetic simulations.
Top-right panel shows the inductive or MHD term (blue 
lines), the Hall term (green lines), electron inertia term (black 
lines). These are shown for both simulations. 
The pressure contribution is the sum of contributions
from the isotropic and the anisotropic part of the pressure
tensor (magenta lines), and can be computed only from the
kinetic simulation. 
The inductive term is the largest 
contribution at the the MHD macro-scales and 
in this range the 
kinetic and two-fluid simulations are in close agreement. 
This is expected since the
physics of the large scale dynamics is well represented 
in both simulation models.

Going beyond the proton scale to smaller scales, 
the inductive term ceases to 
be dominant, and
the Hall term becomes important.
The resulting Hall electric field is similar
in both simulations up to the wavenumber 
corresponding to the electron inertial scale.
Beyond that, additional contributions in Ohm's law are required.

The perpendicular electric field spectrum 
from the electron inertia term shows similar 
spectral shape to the 
inductive term, but at much lower level.
Electron inertial contributions
are somewhat stronger in the kinetic simulation
than in the two-fluid simulation
at sub-proton scales.

The pressure related electric field
effect is small at large scales and plays an 
increasingly important role at sub-proton
scales, and it is comparable to the Hall term 
when approaching the electron inertial length. 
The most relevant contributions to 
the pressure term comes from the non-diagonal terms of the 
pressure tensor which contains the finite electron Larmor radius effects 
that are neglected in the incompressible two-fluid model.
If we do not take into account the pressure effects on the electric field, 
the perpendicular component of the total electric field
for both simulations is very well correlated at all scales (see the top-right
panel or Figure 5 (b)). 

The total perpendicular electric field for both simulations at the large MHD scales
are similar, with a slightly greater perpendicular electric 
field spectral density in the two fluid simulation. 
Moving to scales smaller than the ion inertial 
length, the two models remain similar, but the kinetic simulation has slightly 
greater spectral amplitude that could arise from additional kinetic effects
that are not contained in the Ohm's law terms we computed.
At the electron inertial scale the curves become almost equal, 
and at still smaller scales the spectral density 
in the fluid case is heavily suppressed due to resistivity, while the 
kinetic case indicates and enhancement most likely due to Langmuir 
oscillations.

Figure 5.(c) depicts the spectrum of the parallel electric field.
The most relevant terms of the generalized Ohm's law for
this particular direction are the MHD and Hall terms, the pressure effects
that only contain the contributions from the divergence of pressure tensor because 
the gradient of isotropic pressure cannot be determined in the direction of 
the guide field due to our 2D domain. 
Also the resistive electric field is not shown since it is 
not of physical interest for weakly collisional plasmas, 
and has no counterpart in the collisionless model that we are studying with 
the PIC simulation.

Electron inertia term remains smaller than the Hall term down
to dissipative/Debye scales in the perpendicular field but for parallel electric field, the
inertia term becomes comparable to Hall term just beyond electron scales in PIC. In the 
fluid case, it supersedes the Hall term at scales much larger than the electron scales. 
The pressure effects shows relative importance at sub-proton scales, this parallel electric
field comes from purely kinetic effects due the non-gyrotropic contributions of the 
pressure tensor. Nongyrotropy could play important role in several different phenomena like 
magnetic reconnection, particle acceleration and wave-particle interaction.
Further, the parallel electric field at kinetic scales looks much different for both simulations despite 
the relatively good agreement at large scales. The parallel electric field in PIC 
simulation shows a significant bump between proton and electron scales. This
bump is not explained by any of the terms in Ohm's law and is possibly because
of the non-negiligible contributions from plasma fluctuations at those scales, an
artifact of the unrealistic value of $\omega_{pe}/\omega_{ce}$ in our PIC
simulation. A detailed breakdown of this anomaly is however beyond the scope of present 
paper and is left for future investigations.

The electron inertial term takes a significant contribution 
to the parallel electric field at electron scales but it seems 
those effects are overestimated in the two-fluid model due to the 
dissipative and resistive effects contained in the
calculation of the Ohm's law.

The parallel electric fields would be better studied in the context 
of 3D simulation instead of the present 2D simulation. The contribution of the parallel electric field is 
fundamental to the study of turbulent dissipation in 
plasmas and it seems that we are not correctly resolving the
parallel direction either with the two-fluid or with the 
kinetic simulation.

\section{\label{sec:level5}DISCUSSION:}

In this paper we have studied plasma turbulence at scales 
ranging from macroscopic MHD scales to 
microscopic kinetic scales using two numerical 
models that differ greatly in their physical content.
The plasma descriptions are, 
first, a two-fluid MHD model that contain  Hall and electron inertia effects, and, 
second, a full kinetic plasma description through Particle-in-Cell 
simulation.

The comparison of those two different numerical methods is 
made in the spirit and context of the \textit{``Turbulent dissipation 
challenge''}. Thus,  we constructed 
a comparison of the two physical models with different numerical 
schemes under the same initial conditions with similar physical 
and numerical parameters.

Owing to the inherently distinct nature of the models, 
there are different scales related to each model that the other one cannot
retain. An example is the Debye length in the kinetic 
simulation, which is the smallest scale where the 
quasi-neutrality condition of the plasma is valid. On the other hand, the smallest scale in the fluid 
simulation is the dissipative scale where the energy is 
dissipated by resistive and viscous effects, a scale that is absent in 
PIC model. With this in mind, we compared the 
results in the spatio/temporal scales that both models share in
common, that is, the energy containing scales (related with the 
system size) and the proton and electron scales.
We have explored some features of the turbulence at different scales
and we have found many similarities
in the results, including comparisons 
of overall decay of the energy density,
and the generation
of small-scale magnetic structures mediated by magnetic
reconnection.
We also obtained similar magnetic spectra for both
simulations which are in good agreement with theoretical and data
measurements.

Particularly revealing is the 
study of kinetic features of the electric field 
in the two-fluid and in the
PIC simulation. This affords an assessment of the
relevance of a given effect in the kinetic range. 
We have shown that the perpendicular component to the 
out-of-plane mean magnetic field is dominated by the MHD electric field at the 
macroscale and compares well in both simulations. At the proton scale, the Hall term becomes dominant. This part of the electric field is important for small scale structures
that naturally appear in the simulation, for example, in the formation of plasmoids that arise
in unstable current sheets. The pressure effects 
are relevant at sub-proton scales with significant contributions from the non-diagonal terms of the pressure tensor that contains
the electron finite larmor effects. Those effects are not retained in the
incompressible two-fluid model. 

The comparison of the electric field between these two types of
simulations may help us understand the capacity of the fluid description to retain effects
that appear at kinetic scales. Indeed 
for certain problems the fluid model, that takes into account 
some kinetic effects, may be the optimal choice.  For example, the in-plane electric field is well described by the two-fluid model in some part of the kinetic range although it lacks pressure
gradient effects.

However, the parallel component is not well resolved in the 2D computational
domain  we used here. In addition, the two-fluid simulation 
does not resolve correctly the microphysics in the parallel direction, 
and the enhancement of electric field at kinetic scales revealed in the 
kinetic simulation is not obtained with this model. The
parallel electric field plays a significant role in the energy
dissipation in plasmas as it mediates the wave-particle
interaction through Landau damping and other phenomena that are lacking in
this two-fluid model. Clearly in order to understand 
the whole system 3D simulations are required, which is certainly
expensive and large computational resources must be needed to 
simulate a system size with a representative number of 
particles in the kinetic simulation.
Tiny 3D boxes are not realistic for most 
applications (see e.g., \cite{Parashar2015})
and large 3D PIC simulations are extremely expensive. 
Hence a full comparison between fully 3D kinetic and fluid
simulations is deferred to the future.

The connection of the fluid and kinetic descriptions is relevant 
for understanding plasma dynamics, 
due to the need of more efficient 
tools to resolve the huge
range of scales that exist simultaneously in a plasma.
This paper has provided a modest step 
forward in this direction, providing 
quantitative insights concerning the physics 
retained in two models:
a kinetic simulation, which is more complete and more 
expensive to run, 
and a two-fluid 
simulation which is much less computationally demanding
and which in some cases 
resolves all the scales that are needed. 
It also could open the door 
to a new generation of hybrid codes that contain both fluid
and kinetic model in its base, see for example Bai et al. (2015)\cite{Bai2015}.
 
\section*{Acknowldegments}
C.A.G and P.D. acknowledge support from grants UBACyT No. 20020110200359 and
20020100100315, and from grants PICT No. 2011-1529, 2011-1626, and
2011-0454. TNP was supported by the NSF SHINE grant AGS-1460130.
W.H.M. was partially supported by NASA LWS-TRT grant 
NNX15AB88G, Grand Challenge Research grant NNX14AI63G, and the Solar
Probe Plus mission through the Princeton ISOIS. 
C.A.G acknowledges the support and hospitality
during his visit to University of Delaware.
\nocite{*}
\bibliography{aipsamp_arxiv}

\end{document}